\documentclass[smus]{article}

\begin{document}

\title{ Decays of  Long Lived  Lightest Supersymmetric 
Particles in the Galactic Halo}

\author{ Milind V. Diwan\\
Brookhaven National Laboratory, Upton, NY }

\maketitle

%\begin{center}
%This work was supported by the U.S. Department of Energy under 
%contract\\ DE-AC02-76CH00016.
%\end{center}

\begin{abstract}
If dark matter  neutral LSPs in the galactic halo 
decay into two body final states containing photons or neutrinos
they could be detected even if the decay rates are 
very small, $\sim 10^{-32} s^{-1}$. 
I calculate mass and lifetime bounds from current astrophysical 
data on monochromatic photons and neutrinos and suggest 
that the poorly explored region between 10 GeV and 1 TeV be explored
for signs of supersymmetry from space. 
\end{abstract}

\bigskip 
\medskip

In this note I describe a calculation of the possibility
of detecting two body decays of very long lived 
 lightest supersymmetric 
particles (LSP or $\tilde{\chi}^0_1$).
If the lifetime is  long enough then 
these LSPs could form a significant component of the dark 
matter. It is proposed that the two body decays 
$\tilde{\chi}^0_1 \to \tilde{G} \gamma$, 
$\tilde{\chi}^0_1 \to \tilde{G} \nu$,  $\tilde{\chi}^0_1 \to J \nu$,  
and  $\tilde{\chi}^0_1 \to \nu \gamma$ of the LSPs
 that form the local galactic halo could be 
detected even if the decay rates are very small. 
 $\tilde G$ and $J$ are 
the gravitino and the Majoron, respectively, that 
arise in some SUSY models. For example,
the first two of these 
 decays have been suggested in the context of 
low energy gauge mediated SUSY breaking models [\ref{sthomas}],
though with small lifetimes. The decays to a Majoron and a neutrino 
or a neutrino and a photon
could take place in R-parity violating models in which the violation 
is very small [\ref{valle}].  
Regardless of the theoretical prejudice, if LSPs do form 
majority of the dark matter it must be interesting to 
detect them and to probe their lifetime at the same sensitivity
as proton decay. Such experiments will be important even if 
supersymmetry is detected at present or future colliders because 
the measured density in the galactic halo will 
 provide important information about the early history 
of the universe.

For this calculation I will assume that the LSP has a mass 
of about 50 GeV and  that all of the local halo
consists of LSPs.
The local galactic halo has a density of about $0.2-0.4 {\rm ~GeV/cm^3}$
[\ref{local_halo}]; therefore there are about $\rho=6\times 10^{-3} 
{\rm~cm^{-3}}$
of LSPs around us. As these LSPs decay 
through two body decays 
monochromatic gammas or neutrinos
with energy  $\sim$25 GeV will result. The flux of these particles on 
the earth will be
$$\phi \approx {\rho \over \tau}\times {\lambda (1-e^{-R/\lambda})}  $$
where R is the radius of the local halo around the galaxy,
$\tau$ is the effective lifetime in the two body decay, and $\lambda$ is
the attenuation length of photons through the galaxy. Since 
the galaxy is mostly transparent to photons the formula simplifies 
to $\rho R\over \tau$. This flux will be diffuse or uniform over all 
angles.  
I will assume $R\approx 30 \rm{~kpc}$ or approximately 3 times the distance 
of the sun from the galactic center. Then the flux can be written as
$$\phi = ({50 {\rm~GeV}\over M_{\chi}}) 
\times {5.5\times 10^{20} {\rm ~cm^{-2}}\over \tau}$$ 
where $M_\chi$ is the mass of the supersymmetric particle.

Consider an electro-magnetic calorimeter with a surface 
area of about $10 {\rm ~m} \times 10 {\rm ~m}$ and angular 
acceptance of $\pi/2$ str in orbit around the earth 
then the rate of gamma events per year will be 
$$N_\gamma = {1\over 4} 5.5\times 10^{33}/\tau$$ 
where $\tau$ is the decay lifetime in seconds. 
Thus if we 
require about 10 events in the 25 GeV peak then lifetimes of 
the order of $\sim 10^{32}$ sec could  be reached. 

Currently there is only one experiment that could look at such photons
from outer space [\ref{cosmic_g}]. The EGRET experiment has observed the
diffuse spectrum of gamma rays up to 10 GeV. The diffuse background
spectrum at 10 GeV 
 has been
determined to be $10^{-8} {\rm ~cm^{-2} s^{-1} sr^{-1} GeV^{-1}}$.
This background flux has been explained to be 
mainly from interactions of cosmic ray nucleons 
on the galactic matter with significant components from electron 
bremsstrahlung, inverse Compton, and unresolved point sources (blazars)
[\ref{dingus}].
 The energy resolution of EGRET at 10 GeV is about 1 GeV and the solid
angle acceptance is approximately $0.15\pi$ str [\ref{thompson}]. If
we assume that a diffuse monochromatic flux about 10 times larger than
the background at 10 GeV could be considered a signal for
supersymmetric particles then EGRET is sensitive to a flux of about
$5\times 10^{-8} {\rm ~cm^{-2} s^{-1}}$.  Thus EGRET rules out long lived
LSPs of up to 20 GeV with lifetimes less than $\sim 3\times 10^{28} {\rm~s}$.
Since LSPs up to $\sim$20 GeV are already ruled out by LEP data
[\ref{aleph}] it is desirable to push the mass limit to higher values.
The background from unresolved point sources is clearly a function of
the angular resolution and the size of the detector. A new larger
detector (the GLAST experiment) with much better angular and energy
resolution is being considered [\ref{glast}].  This detector will also
have acceptance to photons up to 300 GeV. Such a detector will indeed
be able to explore much of this physics. If a signal is detected
then its variation across the galactic plane 
could be measured. 
 
We will now consider terrestrial experiments to detect the
 monochromatic neutrinos using upward going muons or energetic
 electrons.  The two largest neutrino detectors with the most amount
 of data so far are Kamioka and IMB.  Neither of these
 detectors can measure the momentum of high energy muons that pass
 through the detector, and so the best limits from these detectors
 come from the electron data.  We will consider the sensitivity of
 Kamioka detector only since the data above 10 GeV is published 
[\ref{fukuda}]. 
 The Kamioka detector has had a total
 exposure of 8.2 kTon-yr including fiducial cuts,
 and the spectrum of contained electrons has 6
 events  between 10 GeV and 50 GeV. These events are
 consistent with background from atmospheric neutrinos.  We will assume 
that a mono-energetic peak of events approximately 10 times the observed 
background  constitutes a signal (about 10 events in the peak).
 Then using the 
charged current neutrino (anti-neutrino) cross section of 
$\sim 0.7\times 10^{-38} E_\nu {\rm  ~cm^2 GeV^{-1}}$ 
($\sim 0.3\times 10^{-38} E_{\bar\nu} {\rm  ~cm^2 GeV^{-1}}$) 
we obtain the following lifetime limits independent of 
the mass of the LSP:
\begin{eqnarray}
\tau(\tilde\chi^0_1 \to X \nu_e) > 1.4 \times 10^{24} {\rm ~s} \\
\tau(\tilde\chi^0_1 \to X \bar\nu_e) > 0.6 \times 10^{24} {\rm ~s} 
\nonumber
\end{eqnarray}
where $X$ is any light neutral particle.
The Super-Kamioka detector, which started operating recently [\ref{superk}],
has 
approximately 5 times the mass of the Kamioka detector; it 
also has much better energy resolution and particle identification,
and so it should be able to reach 10 times longer lifetimes.

It is interesting to note that there is a lack of data about both
neutrinos and photons of astronomical origins between 10 and a few
hundred GeV. Since supersymmetric particles are postulated to have
masses in this range, careful examination of both neutrinos and
photons of a few tens of GeV is quite important.
A cosmic ray muon  detector to  measure the momentum of 
upward going muons could partially fill  this gap [\ref{wonyong}].
As an example consider a detector of size about 
$30{\rm ~m} \times 30 {\rm ~m}$ 
and an angular acceptance of about $\pi/2$ str. 
The detector would be composed of an iron toroid magnet with 
a total momentum kick of about 1 GeV with tracking chambers 
to measure the muon bend. 
Water Cherenkov detectors placed above and below the iron 
toroid would detect the time and the direction of the 
muon.  
Such a detector would 
have an effective fiducial mass greater than 50 kT for 20 GeV 
muon neutrinos. Although it would be difficult to detect a mono-energetic 
peak with such a detector without any knowledge of the event vertex, 
an excess of high energy ($>10 {\rm~GeV}$) muon events 
above the atmospheric neutrino background will clearly signal 
interesting physics. 

I would like to thank  John Womersley, Alfred Mann,  
Sid Kahana, and Robert Harr for useful discussions. 

\bigskip

{\bf REFERENCES}

\begin{enumerate}

%\item \label{age_of_nu}
% J. Bernstein, G. Feinberg, Phys. Lett. {\bf 101B} (1981) 39.
%      ERRATUM-ibid, {\bf 103B} (1981) 470.
%\vspace{-10pt}

\item\label{sthomas}
S. Dimopoulos, M. Dine, S. Raby, S. Thomas,
Phys. Rev. Lett. {bf 76} (1996) 3494. Also see the 
presentation by S. Thomas during this workshop.

\item\label{valle}
V. Berezinsky, A. Masiero, J.W.F. Valle, Phys. Lett. \underline{B266}, 
1991 (382).

\item \label{local_halo}
 E. Gates, G. Gyuk, M. Turner, FERMILAB-PUB-95/090-A.

\item \label{cosmic_g}
 F.W. Stecker (NASA, Goddard). ASTROPH-9607037, Jul 1996. 5pp. Talk given at 2nd Symposium on
Critique of the Sources of Dark Matter in the Universe, Santa Monica, CA, 14-16 Feb 1996.

\item \label{dingus}
B.L. Dingus, Proceedings of the 1994 Snowmass Summer Study 
on Particle and Nuclear Astrophysics and Cosmology in the 
next millennium, Edited E.W. Kolb and R.D. Peccei, 
 World Scintific, Snowmass, Colorado,
June 29-July 14, 1994.

\item \label{thompson}
Thomson, D.J., et al., ApJ Suppl., {\bf 86} (993) 629.

\item \label{aleph}
D. Buskulic, et al. CERN-PPE-96-83, Jul 1996.  

\item \label{glast}
W.B. Atwood for the GLAST collaboration, 
Nucl. Instrum. Meth. {\bf A342} (1994) 302.
Also see Elliott D. Bloom for the GLAST collaboration, SLAC-PUB-95-6738, Oct 1994. 26pp. Talk given at
International Heidelberg Workshop on TeV Gamma-ray Astrophysics, Heidelberg, Germany, 3-7 Oct 1994.

\item \label{fukuda}
Y. Fukuda et al., Phys. Lett. {\bf B335} (1994) 237.

\item \label{superk}
By Y. Totsuka (Tokyo U., ICEPP). UT-ICEPP-86-06, Presented at 7th Workshop on Grand
Unification, ICOBAN '86, Toyama, Japan, Apr 16-18, 1986. 
Published in Toyama ICOBAN 1986:118

\item \label{wonyong}
Such a detector was suggested by Won Yong Lee
to detect annihilations of WIMPs inside the sun, Private communication.

%\item \label{macro}
%M. Ambrosio et al., Phys. Rev. {\bf D52} 1995 (3793).

\vspace{-10pt}

\end{enumerate}

\end{document}